\documentstyle[12pt,epsfig]{article}

\def\beq{\begin{equation}}
\def\eeq{\end{equation}}
\def\beqa{\begin{eqnarray}}
\def\eeqa{\end{eqnarray}} 

\begin{document}
 
\begin{titlepage}

\begin{flushright}
{\sc UMHEP-430}\\ [.2in]
{\sc June 6, 1996}\\ [.5in]
\end{flushright}

\begin{center}

{\LARGE
Form factor relations for heavy-to-heavy and heavy-to-light
meson transitions.}\\ [.5in]
{\large Jo\~{a}o M. Soares}\\ [.1in]
{\small 
Department of Physics and Astronomy, University of Massachusetts,\\
Amherst, MA 01003-4525}\\ [.5in]

{\normalsize\bf 
Abstract}\\ [.2in]

\end{center}

{\small
Relations between the form factors that parametrize the hadronic matrix 
elements, in spectator decays of heavy mesons, have been derived by Stech
within the constituent quark model. In here, we examine these relations using 
a slightly modified description of the meson states.  We find new and very 
general relations for some of the form factors. For the other form factors,
we obtain small modifications to the relations previously derived by Stech,
in the case of heavy-to-light transitions.} 
 
\end{titlepage}

\section{Introduction}

In ref.~\cite{Stech}, Stech derived a set of relations between the form 
factors that parametrize the hadronic matrix elements, in the weak decays of 
heavy mesons. These relations follow from the constituent quark picture for 
the hadronic transitions, whenever spectator effects can be ignored, and they 
are independent of the particular model that is chosen for the wavefunctions
of the mesons. For heavy-to-heavy meson transitions, the relations are 
consistent with those obtained from the Heavy Quark Symmetry limit 
\cite{Neubert}. For heavy-to-light meson transitions, the relations obtained 
by Stech are important new results. For example, they can be useful in 
extracting the CKM matrix elements $|V_{ub}|$ (from the exclusive 
semi-leptonic decays $B \to \pi l \bar{\nu}_l$, $\rho l \bar{\nu}_l$, etc.), 
or $|V_{td}|$ (from the exclusive radiative decay $B \to \rho \gamma$); or 
they can help us understand better the color-suppressed hadronic decays, such 
as the $B$ decays into the charmonium states.

In here, the derivation of the relations between the form factors is 
revisited, along similar steps to those in ref.~\cite{Stech} (however,
we adopt a slightly different description of the kinematics of the 
constituent quarks inside the mesons). We will show that new and very general
relations can be derived for some of the form factors; they follow from the 
constituent quark description and the constraint of a spectator decay, but 
with no further assumptions. Additional relations between the other form
factors are obtained in two limiting cases. For 
heavy-to-heavy transitions, we obtain the same results as Stech. For the 
heavy-to-light case, we analyze in more detail the conditions under which the
approximations already used by Stech are valid. Using our description of the 
constituent kinematics, we confirm the surprising result that the static heavy
quark approximation remains valid, even for decays with large recoil. On the 
other hand, we find that the approximation of a massless recoiling constituent
is valid in a more restricted region of recoil momentum. When both 
approximations are used, the results of Stech are recovered, up to differences
of the order of the ratio of the light to heavy meson masses. 

The derivation is presented in some detail, although part of the results can 
already be found in ref.~\cite{Stech}. For definiteness, we shall refer to the
heavy decaying meson as a $B$ meson, although part of the results may also
apply to $D$ decays.

\section{Form factor relations.}

\subsection{Hadronic matrix elements in the constituent quark model.}

The hadronic matrix elements of interest, in spectator decays of $B$-mesons, 
are of the form 
\beq
\langle X(\vec{p^\prime})|\overline{q} \Gamma b|B(\vec{p})\rangle \ ,  
\hspace{.3in}
\Gamma = \gamma^\mu, \gamma^\mu \gamma_5, 
i \sigma^{\mu\nu} (p - p^\prime)_\nu,
i \sigma^{\mu\nu} (p - p^\prime)_\nu \gamma_5 \ .
\label{10}
\eeq
In here, the meson $X$ will be a pseudoscalar $P$ or a vector meson $V$. It 
can be a heavy meson such as a $D$ or $D^\ast$, or a light meson such as a 
$\pi$ or $\rho$. In the constituent quark model, the $B$ and $X$ mesons are 
the quark-antiquark bound states $(b\overline{q}_{sp})$ and 
$(q\overline{q}_{sp})$, respectively. For a spectator decay, the operator 
$\overline{q} \Gamma b$ annihilates the $b$-quark in the initial state, and 
creates the constituent quark $q$ in the final state. It does not act on  
$\overline{q}_{sp}$, which behaves as a spectator: its momentum and spin remain
unchanged in the decay process.

The $B$ meson state in eq.~\ref{10} is 
\beq
|B(\vec{p})\rangle 
= \sqrt{\frac{2 E}{(2\pi)^3}} \int d^3 \vec{k} \; \phi_B(|\vec{k}|)
\sqrt{\frac{m_{sp}}{E_{sp}}} \sqrt{\frac{m_b}{E_b}} 
\chi^{0 0}_{\overline{\sigma} \sigma} 
|\overline{q} (\vec{p}_{sp},\overline{\sigma}) \rangle
|b(\vec{p}_b,\sigma)\rangle  \ ,
\label{20}
\eeq
and similarly for the $X$ meson state. The momentum wavefunction $\phi_B 
(|\vec{k}|)$ (with $L=0$) describes the distribution in internal momentum 
$\vec{k}$. It favors small values of $|\vec{k}|$ (of the order of a typical
hadronic mass scale, $\Lambda_{QCD}$), and falls off rapidly for large values
of the internal momentum \cite{Normalizations}. The spin wavefunction is 
$\chi^{0 0}_{\overline{\sigma} \sigma} = (|\uparrow\downarrow\rangle -
|\downarrow\uparrow\rangle)/\sqrt{2}$.

The hadronic matrix element of eq.~\ref{10} is then \cite{Normalizations}
\beqa
\lefteqn{\langle X(\vec{p^\prime})|\overline{q} \Gamma b|B(\vec{p})\rangle 
= \sqrt{2E}\sqrt{2E^\prime} \int d^3\vec{k^\prime} \phi_B(|\vec{k}|) 
\phi_X^\ast(|\vec{k^\prime}|)} \nonumber \\
& & \times \sqrt{\frac{m_b}{E_b}} \sqrt{\frac{m_q}{E_q}} 
\chi^{J m_J}_{\sigma^\prime\overline{\sigma}}
\chi^{00}_{\overline{\sigma}\sigma}
\overline{u}_q(\vec{p}_q,\sigma^\prime) \Gamma u_b(\vec{p}_b,\sigma) \ ,
\label{50}
\eeqa
where the condition that the spin of the spectator $\overline{q}_{sp}$ is not 
affected by the decay has been included. The condition that its momentum 
remains unchanged leads to a relation between the internal momenta $\vec{k}$ 
and $\vec{k}^\prime$, respectively of the $B$ and $X$ mesons. For 
definiteness, we work on the $B$ rest frame, $\vec{p}=0$, and choose the 
z-axis along the momentum of the $X$ meson, $\vec{p^\prime} = |\vec{p^\prime}|
\hat{z}$. In the absence of internal momentum, the constituents are at rest in
the meson c.\ m.\ frame; but in general,
\beqa
p_{b,\bot} = k_\bot & & p_{b,z} = k_z 
\label{60} \\
p_{sp,\bot} = -k_\bot & & p_{sp,z} = -k_z 
\label{70}
\eeqa
for the $B$ meson, and 
\beqa
p_{q,\bot} = k^\prime_\bot  & & 
p_{q,z} = \gamma \left( k^\prime_z + \beta \sqrt{m_q^2 + 
\vec{k^\prime}^2} \right) 
\label{80} \\
p_{sp,\bot} = -k^\prime_\bot  & & 
p_{sp,z} = \gamma \left( -k^\prime_z + \beta 
\sqrt{m_{sp}^2 + \vec{k^\prime}^2} \right)
\label{90}
\eeqa
for the $X$ meson, with  $\beta = |\vec{p^\prime}|/E^\prime$ and $\gamma = 
1/\sqrt{1-\beta^2}$. This gives 
\beqa
k_\bot = k^\prime_\bot 
& &
k_z = \frac{E^\prime}{m_X} k^\prime_z -  \frac{|\vec{p^\prime}|}{m_X}
\sqrt{m_{sp}^2 + \vec{k^\prime}^2}  \ , 
\label{150}
\eeqa
which is the relation between the internal momenta $\vec{k}$ and 
$\vec{k}^\prime$, that is implicit in eq.~\ref{50}. 

As in ref.~\cite{Stech}, we assume that the wavefunctions of the $B$ and $X$ 
mesons are strongly peaked (at $|\vec{k}|, |\vec{k^\prime}| \simeq 0$). Then, 
the overlap of the wavefunctions in eq.~\ref{50} will also be strongly peaked;
and so, to a good approximation, 
\beqa
\langle X(\vec{p^\prime})|\overline{q} \Gamma b|B(\vec{p})\rangle &=&
\sqrt{4 m_B E^\prime \frac{m_b}{E_b}\frac{m_q}{E_q}} R_{B \to X} 
f_{\sigma^\prime \sigma} \nonumber \\ 
& & \times 
\overline{u}_q(\vec{p}_q,\sigma^\prime) \Gamma u_b(\vec{p}_b,\sigma) \ ,
\label{160}
\eeqa
with $R_{B \to X} \equiv \int d^3\vec{k^\prime} \phi_B(|\vec{k}|) 
\phi_X^\ast(|\vec{k^\prime}|)$ and $f_{\sigma^\prime\sigma} \equiv 
\chi^{J m_J}_{\sigma^\prime\overline{\sigma}} 
\chi^{00}_{\overline{\sigma}\sigma}$. 
This is the expression for the hadronic matrix elements that we will use to
derive the form factor relations, in the next section.

In order to determine the range of recoil momentum $|\vec{p^\prime}|$ where
the form factor relations are valid, we will need to compare the quark momenta 
$|\vec{p}_b|$ and $|\vec{p}_q|$ with the corresponding quark masses $m_b$ and 
$m_q$. The quark momenta in eq.~\ref{160},
\beqa
p_{b,\bot} = \underline{k}_\bot & & p_{b,z} = \underline{k}_z 
\label{190} \\
p_{q,\bot} = \underline{k}_\bot & & 
p_{q,z} = |\vec{p^\prime}| + \underline{k}_z  \ , 
\label{200}
\eeqa
are determined by the location $\vec{\underline{k}}$ of the peak in the 
overlap of the meson wavefunctions. The quark masses $m_b$ and $m_q$ are 
effective masses determined by the relations
\beqa
m_B &=& \sqrt{m_b^2 + \vec{\underline{k}}^2} + 
\sqrt{m_{sp}^2 + \vec{\underline{k}}^2} 
\label{208} \\
m_X &=& \sqrt{m_q^2 + \vec{\underline{k}^\prime}^2} + 
\sqrt{m_{sp}^2 + \vec{\underline{k}^\prime}^2} \ ,
\label{209}
\eeqa
and they too depend on $\vec{\underline{k}}$ (the corresponding value of 
$\vec{\underline{k}^\prime}$ is given by the relations in eq.~\ref{150}).
On the other hand, $m_{sp}$ is a constant, and a parameter of the model; its 
maximum allowed value is determined by eqs.~\ref{208}--\ref{209}. 

The transverse component of the internal momentum $\vec{\underline{k}}$ is
$\underline{k}_\bot \simeq 0$, whereas the longitudinal component 
$\underline{k}_z$ is more sensitive to the relative shape of the $B$ and $X$ 
wavefunctions. For $B$ and $X$ wavefunctions that are similar (scenario 
{\bf A}), both mesons share the internal momentum that is required for the 
spectator transition. Then,
\beq
{\rm {\bf A}}:\hspace{.3in}
\underline{k}_z \simeq - \frac{1}{2} 
m_{sp} \frac{|\vec{p^\prime}|}{m_X} 
\sqrt{\frac{2m_X}{E^\prime+m_X}} \ .
\label{205}
\eeq
If the $B$ wavefunction has a much narrower spread in internal momentum
(scenario {\bf B}), then 
\beq
{\rm {\bf B}}:\hspace{.3in}
\underline{k}_z \simeq 0 \ .
\label{206}
\eeq
If, on the contrary, the $X$ wavefunction is much narrower (scenario 
{\bf C}), then
\beq
{\rm {\bf C}}:\hspace{.3in}
\underline{k}_z \simeq - m_{sp} \frac{|\vec{p^\prime}|}{m_X} \ .
\label{207}
\eeq

\subsection{Form factor relations.}

The hadronic matrix elements in eq.~\ref{10} are parametrized in terms of 
Lorentz invariant form factors as follows:
\beqa 
\langle P(\vec{p^\prime})|\overline{q}\gamma^\mu b|B(\vec{p})\rangle
&=& (p + p^\prime)^\mu f_1(q^2) \nonumber \\
& & + \frac{m_B^2 - m_P^2}{q^2} q^\mu [ f_0(q^2) - f_1(q^2) ] \ ,
\label{210} 
\eeqa 
where $f_1(0) = f_0(0)$;
\beqa 
\langle P(\vec{p^\prime})| \overline{q} i \sigma^{\mu\nu} q_\nu b 
|B(\vec{p})\rangle &=& 
s(q^2) [ (p + p^\prime)^\mu q^2 - (m_B^2 - m_P^2) q^\mu ] \ ; 
\label{220} \\
\langle V(\vec{p^\prime},\vec{\varepsilon})|\overline{q}\gamma^\mu b
|B(\vec{p})\rangle 
&=& \frac{-1}{m_B + m_V} 2 i \epsilon^{\mu\alpha\beta\gamma}  
\varepsilon_\alpha^\ast p_\beta^\prime p_\gamma V(q^2) \ ; 
\label{230} \\
\langle V(\vec{p^\prime},\vec{\varepsilon})|\overline{q}\gamma^\mu \gamma_5
b|B(\vec{p})\rangle
&=& (m_B + m_V) \varepsilon^{\mu\ast} A_1(q^2)  \nonumber \\ 
& & -\frac{\varepsilon^\ast.q}{m_B + m_V} (p+p^\prime)^\mu A_2(q^2)  
\nonumber \\ 
& & - 2 m_V \frac{\varepsilon^\ast.q}{q^2} q^\mu 
[ A_3(q^2) - A_0(q^2) ] \ ,
\label{240} 
\eeqa
where $2 m_V A_3(q^2) \equiv  (m_B + m_V) A_1(q^2) 
- (m_B - m_V) A_2(q^2)$ and $A_0(0) = A_3(0)$;
\beqa 
\langle V(\vec{p^\prime},\vec{\varepsilon})| \overline{q} i  
\sigma^{\mu\nu} q_\nu b|B(\vec{p})\rangle &=& 
i \epsilon^{\mu\alpha\beta\gamma}
\varepsilon_\alpha^\ast p_\beta^\prime p_\gamma F_1(q^2) \ ; 
\label{260} \\
\langle V(\vec{p^\prime},\vec{\varepsilon})| \overline{q} i  
\sigma^{\mu\nu} q_\nu \gamma_5 b|B(\vec{p})\rangle &=&
[(m_B^2 - m^2_V) \varepsilon^{\mu\ast}  
- \varepsilon^\ast.q (p+p^\prime)^\mu] F_2(q^2) \nonumber \\ 
& & + \varepsilon^\ast.q [q^\mu - \frac{q^2}{m_B^2 - m^2_V}  
(p+p^\prime)^\mu] F_3(q^2) \ , \nonumber \\
& &
\label{270}
\end{eqnarray} 
where $F_1(0) = 2 F_2(0)$. In all of the above, $q \equiv p-p^\prime$.

From the result in eq.~\ref{160}, we obtain for each one of the form factors:
\beqa 
f_1(q^2) &=& (\frac{m_B - E^\prime}{|\vec{p^\prime}|} P_+ + Q_+)  
N R_{B \to P}\ , 
\label{280} \\
f_0(q^2) &=& 
\left[ (\frac{m_B - E^\prime}{|\vec{p^\prime}|} P_+ + Q_+) \right. \nonumber\\
& &  \left. - \frac{q^2}{m_B^2 - m_P^2} 
(\frac{m_B + E^\prime}{|\vec{p^\prime}|} P_+ - Q_+) \right] 
N R_{B \to P} \ , 
\label{290} \\
s(q^2) &=& \frac{1}{|\vec{p^\prime}|} P_- N R_{B \to P} \ , 
\label{300} \\
V(q^2) &=& - \frac{m_B + m_V}{|\vec{p^\prime}|} P_- N R_{B \to V}  \ , 
\label{310} \\
A_0(q^2) &=& (\frac{m_B - E^\prime}{|\vec{p^\prime}|} P_+ + Q_+) 
N R_{B \to V} \ , 
\label{320} \\
A_1(q^2) &=& \frac{2m_B}{m_B + m_V} Q_- N R_{B \to V}  \ , 
\label{330} \\
A_2(q^2) &=& \frac{m_B + m_V}{m_B|\vec{p^\prime}|}   
\left[ \frac{m_B E^\prime - m_V^2}
{|\vec{p^\prime}|} Q_- \right. \nonumber \\
& & \left. - m_V ( P_+ + \frac{m_B - E^\prime}{|\vec{p^\prime}|} Q_+) \right]
N R_{B \to V} \ , 
\label{340} \\
F_1(q^2) &=& 
( \frac{m_B - E^\prime}{|\vec{p^\prime}|} P_+ + Q_+ ) 
2 N R_{B \to V}  \ , 
\label{350} \\
F_2(q^2) &=&\frac{2m_B|\vec{p^\prime}|}{m_B^2 - m_V^2}
(P_+ + \frac{m_B - E^\prime}{|\vec{p^\prime}|} Q_+) N R_{B \to V}  \ , 
\label{360} \\
F_3(q^2) &=& \left[ - \frac{m_V(m_B^2 - m_V^2)}{m_B|\vec{p^\prime}|^2} 
Q_-  \right. \nonumber \\
& & \left. + \frac{m_B E^\prime + m_V^2}{m_B|\vec{p^\prime}|} 
(P_+ + \frac{m_B - E^\prime}{|\vec{p^\prime}|} Q_+) \right] N R_{B \to V} \ ,
\label{370}
\eeqa
with
\beqa 
Q_\pm &\equiv& 1 \pm \frac{p_{q,z}p_{b,z}}{(E_q+m_q)(E_b+m_b)} \ , 
\label{380} \\
P_\pm &\equiv& \frac{p_{b,z}}{E_b+m_b} \pm \frac{p_{q,z}}{E_q+m_q} \ . 
\label{390}
\eeqa
The factor $N$ is defined by
\beq
N \equiv \sqrt{\frac{E^\prime}{m_B} \frac{E_q+m_q}{2E_q}  
\frac{E_b+m_b}{2E_b}} \ .
\label{420}
\eeq

The overlap integrals $R_{B \to X}$, as well as the values of the quark
momenta $p_{b,z}$ and $p_{q,z}$ that appear in $Q_\pm$, $P_\pm$ and $N$, 
depend on the detailed shape of the $B$ and $X$ wavefunctions. We are 
interested in deriving relations between form factors that are independent 
of these quantities, and thus free of the model dependence associated with 
them. From eqs.~\ref{280}--\ref{370}, we have
\beqa
F_1(q^2) &=& 2 A_0(q^2) \ ,
\label{421} \\
F_2(q^2) &=& \frac{m_B |\vec{p^\prime}|}{m_V (m_B - m_V)} \left[ 
\frac{m_B E^\prime - m_V^2}{m_B |\vec{p^\prime}|} A_1(q^2) \right.
\nonumber \\
& & - \left. \frac{2 m_B |\vec{p^\prime}|}{(m_B + m_V)^2} A_2(q^2) \right] \ ,
\label{422} \\
F_3(q^2) &=& \frac{m_B + m_V}{2m_V} A_1(q^2)
- \frac{m_B E^\prime + m_V^2}{m_V (m_B + m_V)} A_2(q^2) \ .
\nonumber \\
\label{423}
\eeqa
These relations follow from our initial assumption that spectator effects can 
be neglected, but no other approximations are necessary.

We now analize the two special cases of heavy-to-heavy and heavy-to-light
transitions; approximations for the $b$ and $q$ quark momenta will lead to
additional model independent relations between the form factors.

\subsection{Heavy-to-heavy transitions.}

When $X$ is a heavy meson such as a $D$ or $D^\ast$, the internal momenta 
$|\vec{\underline{k}}|$ and $|\vec{\underline{k}^\prime}|$, in the $B$ 
and $X$ mesons, are at most of order $m_{sp}\simeq \Lambda_{QCD} << m_B,
m_X$. This is true for any of the scenarios {\bf A}, {\bf B} or {\bf C},
and for the entire range of the recoil momentum: $|\vec{p^\prime}| = 0$ --
$(m_B^2 - m_X^2)/2m_B$ or, equivalently, $q^2 = 0$ -- $(m_B - m_X)^2$. Then,
\beqa
|\vec{p}_b| << m_b & & m_b \simeq m_B 
\label{430} \\
|\vec{p}_q| \simeq |\vec{p^\prime}| & & m_q \simeq m_X \ ,
\label{440}
\eeqa
and so, $Q_{\pm}=1$, $P_{\pm} = \pm |\vec{p^\prime}|/(E^\prime + m_X)$. Using 
these results in the expressions for the form factors gives the additional 
relations: 
\beqa
f_1(q^2) &=&  f_0(q^2) \left[1 - \frac{q^2}{(m_B + m_P)^2}\right]^{-1} 
\label{450} \\
&=& - s(q^2) (m_B + m_P) \ , 
\label{460} \\
V(q^2) &=& A_0(q^2)
\label{470} \\ 
&=& A_2(q^2)  
\label{472} \\
&=& A_1(q^2) \left[1- \frac{q^2}{(m_B + m_V)^2} \right]^{-1}  
\ .
\label{490}
\eeqa
These, together with the results in eqs.~\ref{421}--\ref{423}, are the same 
results that follow from the Heavy Quark Symmetry limit
\cite{Neubert}\cite{IsgurWise}. In that limit, the spin symmetry gives $m_P = 
m_V$ and so $R_{B \to P} = R_{B \to V}$ (since the $P$ and $V$ wavefunctions 
are then identical); then
\beq
f_1(q^2) = V(q^2) \ .
\label{500}
\eeq
Moreover, the flavor symmetry gives $R_{B \to X}=1$ for zero recoil (since the
$B$ and $X$ wavefunctions are identical), and so 
\beq
f_1(q^2_{max.}) = V(q^2_{max.}) = \frac{m_B + m_X}{2\sqrt{m_B m_X}} \ .
\label{510}
\eeq
Both results are well-known consequences of the Heavy Quark Symmetry, that
can be derived independently of the constituent quark picture 
\cite{Neubert}.

\subsection{Heavy-to-light transition.}

When $X$ is a light meson, such as a $\pi$ or a $\rho$, there are two 
approximations that lead to model independent relations between form factors. 
The first is that of a static $b$ quark, i.\ e.\ $|\vec{p}_b| << m_b$. In that 
case, $Q_\pm = 1$ and $P_+ = - P_-$, which gives
\beqa
2 m_B s(q^2) &=& - f_1(q^2) + \frac{m_B^2 - m_P^2}{q^2}
[f_0(q^2) - f_1(q^2)] \ ,
\label{511} \\
V(q^2) &=& - \frac{(m_B + m_V)^2}{2m_B(m_B-E^\prime)} A_1(q^2)  
+ \frac{m_B + m_V}{m_B-E^\prime}  A_0(q^2) \ ,
\label{512} \\
&=&  \frac{(m_B + m_V)^3}{2m_B m_V(E^\prime+m_V)} A_1(q^2)  
- \frac{m_B}{m_V}  A_2(q^2) \ .
\label{513}
\eeqa
We have found, as in ref.~\cite{Stech}, that a static $b$ quark is a very good
approximation over the entire range of the recoil momentum. The only exception
is in the extreme case of scenario {\bf C}, and for large recoil. There, the 
validity of the approximation depends significantly on the choice of $m_{sp}$.
In fig.~1, we show the value of $|\vec{p}_b|/m_b$, for the $B \to \pi$ 
(fig.~1a) and the $B \to \rho$ (fig.~1b) transitions, in scenario {\bf A}.
The two curves correspond to the maximum and minimum values for the parameter
$m_{sp}$.

\begin{figure}[htbp]
\unitlength1mm
\begin{picture}(160,140)
\put(20,70){\makebox(85,65)
{\epsfig{figure=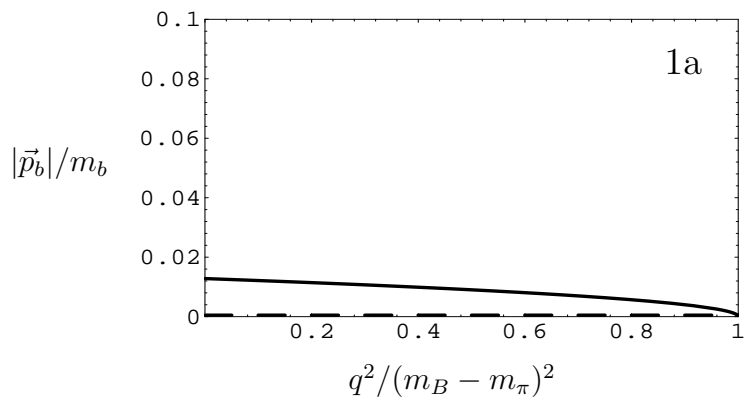,height=100mm,width=80mm}}}
\put(5,103){$|\vec{p}_b|/m_b$}                              
\put(50,74){$q^2/(m_B - m_\pi)^2$}
\put(92,116){\large 1a}
\put(20,0){\makebox(85,65)
{\epsfig{figure=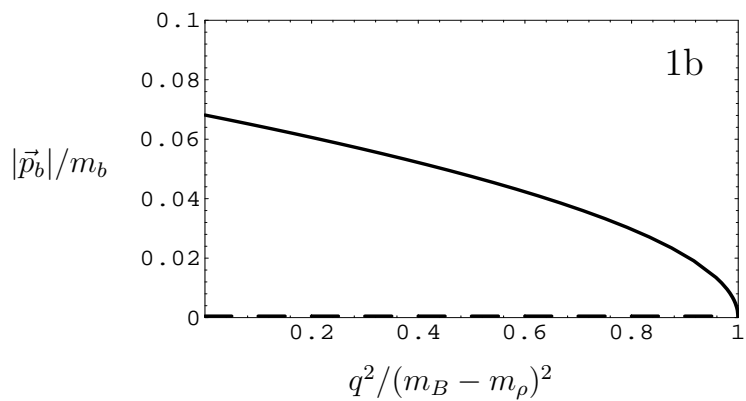,height=100mm,width=80mm}}}
\put(5,33){$|\vec{p}_b|/m_b$}                              
\put(50,4){$q^2/(m_B - m_\rho)^2$}
\put(92,46){\large 1b}
\end{picture}
\caption{Momentum v. mass of the $b$ quark in $B \to \pi$ (fig. 1a)
and $B \to \rho$ (fig. 1b), for $m_{sp}$ maximum (full line) and 
minimum (dashed line).}
\label{fig:1}
\end{figure}

The second approximation is that of a nearly massless recoiling quark, i.\ e.\
$|\vec{p}_q| >> m_q$. Then, $Q_\pm = \pm P_\pm$, and so
\beqa
f_1(q^2) &=& f_0(q^2) \left( 1 - \frac{q^2}{m_B^2 - m_P^2}
\frac{m_B+E^\prime-|\vec{p^\prime}|}
{m_B-E^\prime+|\vec{p^\prime}|} \right)^{-1} \ ,
\label{520} \\
V(q^2) &=& \frac{(m_B + m_V)^2}{2m_B|\vec{p^\prime}|}  A_1(q^2)  
\label{525} \\
&=& \frac{m_B|\vec{p^\prime}|}{m_B E^\prime - m_V^2} A_2(q^2)
+ \frac{m_V(m_B + m_V)}{m_B E^\prime - m_V^2} A_0(q^2)  \  .
\label{530}
\eeqa
The region of validity of this approximation is more restricted than that
of a static $b$ quark: the approximation is valid at high recoil, but it 
breaks down close to lowest recoil; it is worse (in particular for scenario
{\bf C}) for a lower choice of $m_{sp}$. For a very light meson, such as 
$X=\pi$, the validity region expands to most of the kinematic range. In fig.~2,
we show the value of  $m_q/|\vec{p}_q|$, for the $B \to \pi$ (fig.~2a) and the
$B \to \rho$ (fig.~2b) transitions, in scenario {\bf A}. Again, the two curves
correspond to the maximum and minimum values for the parameter $m_{sp}$.

\begin{figure}[htbp]
\unitlength1mm
\begin{picture}(160,140)
\put(20,70){\makebox(85,65)
{\epsfig{figure=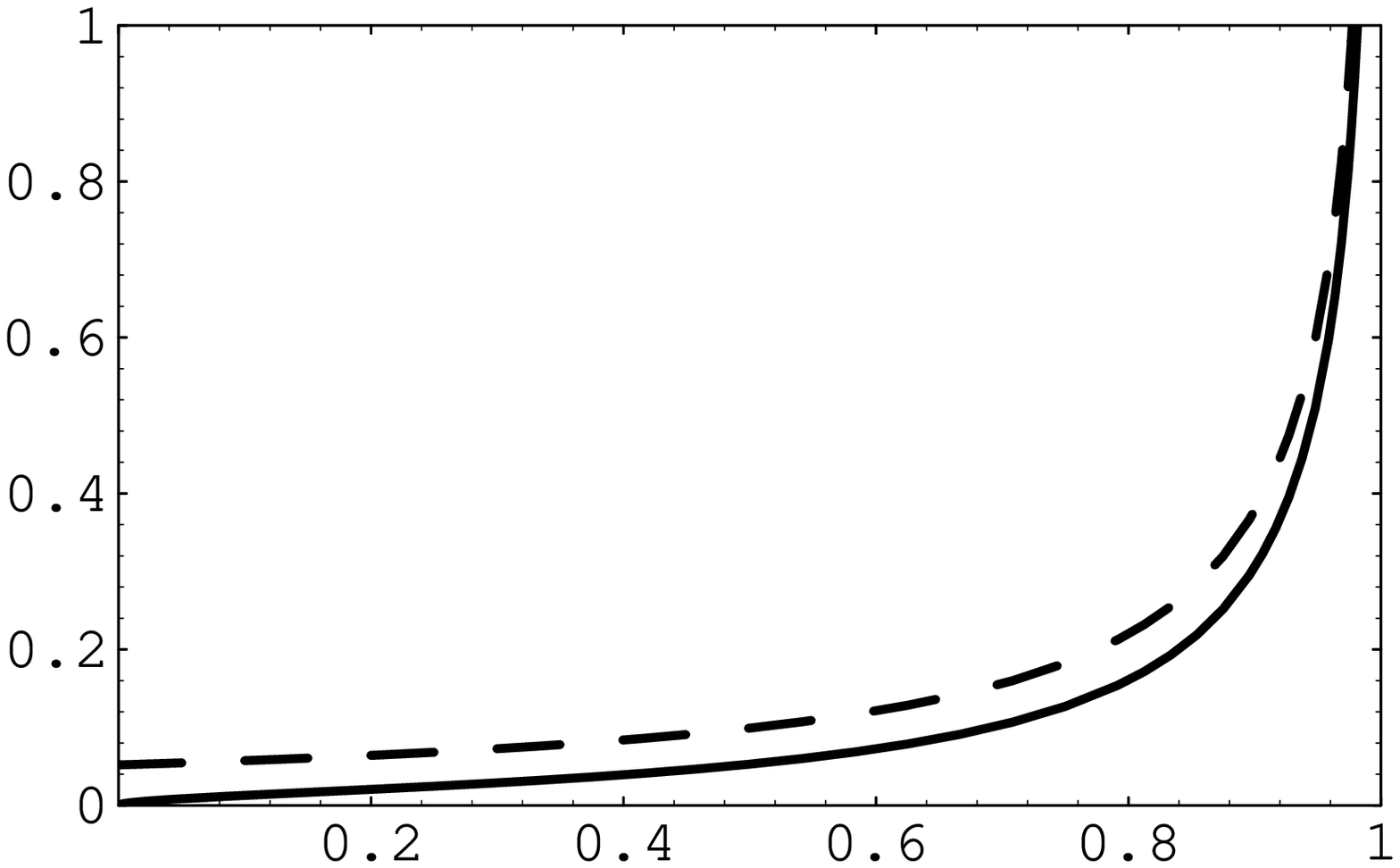,height=100mm,width=80mm}}}
\put(5,103){$m_q/|\vec{p}_q|$}                              
\put(50,74){$q^2/(m_B - m_\pi)^2$}
\put(30,116){\large 2a}
\put(20,0){\makebox(85,65)
{\epsfig{figure=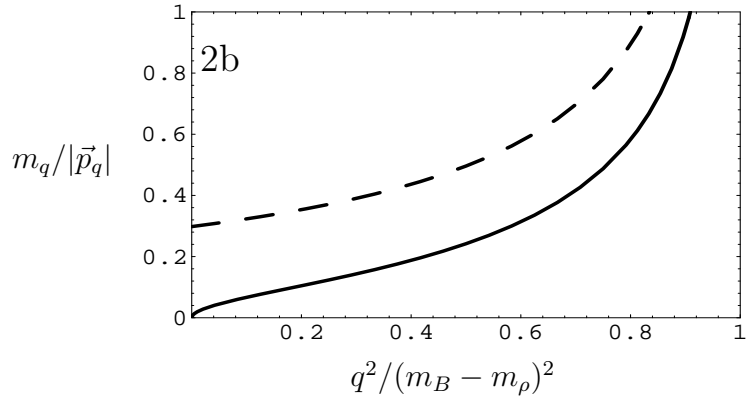,height=100mm,width=80mm}}}
\put(5,33){$m_q/|\vec{p}_q|$}                              
\put(50,4){$q^2/(m_B - m_\rho)^2$}
\put(30,46){\large 2b}
\end{picture}
\caption{Mass v. momentum of the $q$ quark in  $B \to \pi$ (fig. 2a)
and $B \to \rho$ (fig. 2b), for $m_{sp}$ maximum (full line) and 
minimum (dashed line).}
\label{fig:2}
\end{figure}

We have shown separately the consequences of the two approximations, given
that they have different regions of validity. The overlap between the two 
regions is however significant, and in there we have
\beqa
f_1(q^2) &=& f_0(q^2) \left( 1 - \frac{q^2}{m_B^2 - m_P^2}
\frac{m_B+E^\prime-|\vec{p^\prime}|}
{m_B-E^\prime+|\vec{p^\prime}|} \right)^{-1} 
\label{535} \\
&=& - (m_B-E^\prime+|\vec{p^\prime}|) s(q^2) \ ,
\label{540} \\
V(q^2) &=& \frac{(m_B + m_V)^2}{2m_B|\vec{p^\prime}|}  A_1(q^2)  
\label{545} \\
&=&\frac{m_B+m_V}{m_B-E^\prime+|\vec{p^\prime}|}  A_0(q^2)
\label{550} \\
&=& m_B|\vec{p^\prime}| \left[E^\prime (m_B+m_V) -
m_V(m_B + m_V +|\vec{p^\prime}|) \right]^{-1} A_2(q^2)  \  .
\nonumber \\
& & \label{555}
\eeqa
These reproduce the heavy-to-light relations of ref.~\cite{Stech}, up to 
differences in the terms of order $m_X/m_B$ \cite{test}.

\section{Conclusion}

The form factor relations in the constituent quark model, that were proposed 
by Stech in ref.~\cite{Stech}, were looked into in more detail, and using a 
slightly different description of the meson states. We found that the form 
factors for the $\langle V|\overline{q}i \sigma^{\mu\nu} q_\nu b|B \rangle$ 
and $\langle V|\overline{q} i \sigma^{\mu\nu} q_\nu \gamma_5 b|B \rangle$ 
matrix elements are related to the form factors for the $\langle V|
\overline{q}\gamma^\mu \gamma_5  b|B \rangle$ matrix element (see 
eqs.~\ref{421}--\ref{423}). These relations are independent of any 
assumptions, other than that no significant spectator effects are at play
in the hadronic transition. Such relations can be used to study exclusive 
radiative decays, such as $B \to K^\ast \gamma$, but also semi-leptonic decays
like $B \to K^\ast e^+ e^-$, or color suppressed hadronic decays such as $B 
\to J/\psi K^\ast$.

The form factor relations for heavy-to-heavy transitions, in eqs.~\ref{450}
--\ref{490}, are those expected from the Heavy Quark Symmetry limit, and they 
were also found in ref.~\cite{Stech}. In the heavy-to-light case, we derived 
form factor relations that do not depend on the specific details of the meson 
wavefunctions. This is done with either one of two distinct approximations: 
that of a static $b$ quark (see eqs.~\ref{511}--\ref{513}), and that of a 
nearly massless recoiling quark $q$ (see eqs.~\ref{520}--\ref{530}). In each 
case we have discussed the regions where the approximations are expected to 
hold. In particular, we have confirmed the result of ref.~\cite{Stech}, that
the static $b$ quark approximation remains valid throughout the entire 
kinematic range. For the case where both approximations are valid (see 
eqs.~\ref{535}--\ref{555}), our results differ from those of 
ref.~\cite{Stech}, by terms of order $m_X/m_B$.

\section*{}

This work was supported in part by a grant from the National Science 
Foundation.


\begin{thebibliography}{Abcd:efgh}
\bibitem{Normalizations}
The momentum wavefunction is normalized as
\beq
\int d^3 \vec{k} \; |\phi_B(|\vec{k}|)|^2 = 1 \ .
\label{605}
\eeq
The normalization of the meson and quark or antiquark states, as well as the
normalization of the Dirac spinors, is as in ref.~\cite{Itzykson}.
\bibitem{test}
As a simple check of the heavy-to-light form factor relations, in the limit
$|\vec{p}_q| >> m_q$, we can consider the hadronic matrix elements
\beqa
A^{(\gamma)}_\pm &\equiv& \langle V(\lambda=\pm 1)| 
\bar{q}\gamma^\mu(1-\gamma_5)b |B \rangle 
\varepsilon_\mu^\ast \nonumber \\
&=& (m_B + m_V) \left[ - A_1(q^2) \pm V(q^2) 
\frac{2 m_B |\vec{p^\prime}|}{(m_B + m_V)^2} \right]
\label{610}
\eeqa
and
\beqa
A^{(\sigma)}_\pm &\equiv& \langle V(\lambda=\pm 1)| 
\bar{q}i\sigma^{\mu\nu}q_\nu(1+\gamma_5)b 
|B \rangle \varepsilon_\mu^\ast \nonumber\\ 
&=&  (m_B^2 - m_V^2) \left[ F_2(q^2) \mp F_1(q^2) 
\frac{m_B |\vec{p^\prime}|}{m_B^2 - m_V^2} \right] \ ,
\label{620}
\eeqa
for a $B$ decay into a light vector meson $V$ and a vector particle with 
polarization $\varepsilon_\mu$, with helicities $\lambda=\pm 1$. The operators 
$\bar{q}\gamma^\mu(1-\gamma_5)b$ and $\bar{q}i\sigma^{\mu\nu}q_\nu(1+\gamma_5)
b$ create a left-handed quark $q$. In the limit $m_q \to 0$, this corresponds 
to a quark with helicity $\lambda_q = -1/2$, and so the transition to a vector
meson $V$ with $\lambda = +1$ is not possible, if no spectator effects are 
allowed. Using the form factor relations for the heavy-to-light case, we can 
see immediately that $A^{(\gamma,\sigma)}_+=0$, as expected.
\end{thebibliography}
\end{document}